 \documentclass[aps,pre,showpacs,floats, twocolumn,superscriptaddress,longbibliography,showkeys]{revtex4-1}
\usepackage{lipsum}
\usepackage{mathrsfs}
\usepackage{bm,amsbsy,amssymb,amsmath}
\usepackage{caption}
\usepackage{subcaption}
\usepackage{graphics,graphicx,dcolumn,fleqn,epic,eepic,float,tabularx}
\usepackage{multirow,rotate,rotating,color}
\usepackage[utf8]{inputenc}
\newcommand{\figref}[1]{Fig.~\ref{fig:#1}}
\newcommand{\eqnref}[1]{Eq.~(\ref{eq:#1})} 
  \definecolor{tuered}{RGB}{214,0,74}
  \definecolor{tueblue}{RGB}{0,102,204}
  \newcommand{\revisedtext}[1]{\textcolor{black}{#1}}

\usepackage{tikz,pgfplots}
\usepackage{relsize}
\tikzset{fontscale/.style = {font=\relsize{#1}}}
\usetikzlibrary{calc}
\graphicspath{{img/}}

\begin{document}
\title{Simulations of inertial liquid-lens coalescence
with \\ the pseudopotential lattice Boltzmann method}
  \author{Qingguang Xie}
   \email{q.xie@fz-juelich.de}
    \affiliation{Helmholtz Institute Erlangen-N\"urnberg for Renewable Energy (IET-2), Forschungszentrum J\"ulich, Cauerstra{\ss}e 1, 91058 Erlangen, Germany}
   \author{Jens Harting}
   \email{j.harting@fz-juelich.de}
   \affiliation{Helmholtz Institute Erlangen-N\"urnberg for Renewable Energy (IET-2), Forschungszentrum J\"ulich, Cauerstra{\ss}e 1, 91058 Erlangen, Germany}
\affiliation{Department of Chemical and Biological Engineering and Department of Physics, Friedrich-Alexander-Universit\"at Erlangen-N\"urnberg, Cauerstr.\,1, 91058 Erlangen, Germany}
\date{\today}
\begin{abstract}
The coalescence of liquid lenses is relevant in various applications, including inkjet printing and fog harvesting.
However, the dynamics of liquid-lens coalescence have been relatively underexplored, particularly in the case of liquid lenses with larger contact angles.
We numerically investigate the coalescence of low-viscosity liquid lenses by means of the pseudopotential multi-component lattice Boltzmann method over a wide range of contact angles. In two-dimensional simulations, our numerical results on the growth of the bridge height are in quantitative agreement with experimental measurements for small contact angles. 
In addition, by comparing our simulation results with a theoretical approach based on the thin-sheet equations for liquid lenses, we find that the thin-sheet equations accurately capture the bridge-growth dynamics up to contact angles of approximately $\theta < 40^{\circ}$.
For the three-dimensional case, the growth of the bridge radius is independent of the equilibrium contact angle of the liquid lenses at the initial stage of growth. The dependency between the growth of the bridge height and the bridge radius exhibits a non-linear to linear transition.
\end{abstract}


 %\pacs{
%Coalescence, liquid lenses, lattice Boltzmann method
%}
\maketitle

\section{Introduction}
The coalescence of droplets is a fundamental process relevant to natural phenomena 
and many industrial applications, such as formation of rain drops~\cite{Low1982},  stability of emulsions~\cite{Goff1997,Kumar1996}, enhanced oil recovery~\cite{Perazzo2018}, coating~\cite{Eslamian2018} and printing~\cite{Wijshoff2018}. 
Coalescence is initiated when two droplets come into contact and form a growing liquid bridge. Finally, the two droplets merge into a single droplet, which relaxes to its equilibrium shape. 
The growth of the bridge is controlled by the interplay of capillarity, viscosity and inertia, and exhibits three different dynamic regimes: the inertial limited viscous regime~\cite{Paulsen2012,Paulsen2013,Anthony2020}, the viscous regime~\cite{Hopper1984,Eggers1999,Hopper1990}, and the inertial regime~\cite{Aarts2005, Paulsen2011}. 
\revisedtext{These regimes are typically characterized by the Ohnesorge number, $Oh = \eta / \sqrt{\rho R \gamma}$, where $\eta$ is the dynamic viscosity, $\rho$ is the droplet liquid density, $R$ is the droplet radius, and $\gamma$ is the surface tension. This dimensionless number 
measures the dominance of viscosity over inertia. 
For $Oh \gg 1$, coalescence occurs in the viscous regime, whereas $Oh \ll 1$ indicates the inertial regime. The corresponding characteristic timescales are the viscous time, $\tau_v = \eta R / \gamma$, and the inertial time, $\tau_i = \sqrt{\rho R^3 / \gamma}$~\cite{Jacco2025}. }
The coalescence of suspended droplets in a fluid phase~\cite{Eggers1999, Duchemin2003, Aarts2005, Thoroddsen2007, Paulsen2011,Anthony2020} and sessile droplets on a substrate~\cite{Ristenpart2006, Narhe2008, Lee2012, HernandezSanchez2012, Eddi2013, Stone2022} have been investigated extensively.
At the initial stage of coalescence, in the viscous regime, the radius of the bridge grows as $r_0 \sim t$ for both spherical droplets and sessile droplets~\cite{HernandezSanchez2012},  whereas, in the inertial regime, the radius of bridge grows following $r_0 \sim t^{1/2}$ for spherical droplets~\cite{Burton2007,Paulsen2011, Xia2019} and $r_0 \sim t^{2/3}$ for sessile droplets on a substrate~\cite{Eddi2013}, which indicates that the geometry of droplets has a strong effect on the coalescence dynamics. The crossover from the viscous regime to the inertial regime of the bridge growth for spherical drops depends on fluid properties and drop size~\cite{Paulsen2011, Xia2019}.

However, the coalescence of liquid lenses at a liquid layer has been less investigated~\cite{Jacco2025}, despite its significance in applications such as water harvesting, wet-on-wet printing and droplets on a lubricated surface. Burton et al.~\cite{Burton2007} found that the bridge radius $r_0$ of coalescing dodecane lenses at a water-air interface grows following a $t^{1/2}$ scaling law in the inertial regime, surprisingly the same as that of suspended droplets. Recently, Hack et al.~\cite{hack2020} investigated the coalescence of dodecane lenses at a water-air interface experimentally and applied a theoretical approach based on the thin-sheet equations to describe the growth of the bridge height. They found that the growth of the vertical bridge height $h_0$ follows a $h_0 \sim t$ scaling in the viscous regime and a $h_0 \sim t^{2/3}$ scaling in the inertial regime, which agrees with the theoretical prediction. Very recently, Scheel et al.~\cite{Scheel2023} performed color-gradient lattice Boltzmann (CGLB) simulations of the coalescence of liquid lenses over a wide range of surface tension and viscosity values, successfully capturing the asymptotic temporal behavior in both the viscous and inertial limits. Furthermore, Padhan et al.~\cite{Padhan2023} applied a three-phase Cahn–Hilliard–Navier–Stokes (CHNS) approach to study the spatiotemporal evolution of the fluid velocity, vorticity, and the concentration fields in liquid-lens coalescence.  The above experimental and numerical studies are primarily limited to liquid lenses with small contact angles and only address the initial stage of coalescence, leaving the coalescence of liquid lenses with large contact angles and the coalescence dynamics at later stages largely unexplored.
 
Here, we numerically investigate the coalescence of two low-viscosity liquid lenses at a liquid-liquid interface. We apply the pseudopotential multi-component lattice Boltzmann (PMLB) method to simulate the fluid phases in two and three dimensions. The PMLB method may offer a significant reduction in computational cost and a simpler implementation compared to both the CGLB and CHNS approaches, but leaves less flexibility in, for example, the choice of surface tensions.
We investigate the impact of contact angles on the initial coalescence dynamics of liquid lenses and compare simulation results with experimental findings and theoretical analysis based on the thin-sheet equations. In the two-dimensional simulations, our results agree quantitatively with experimental results and theoretical analysis for small contact angles, $\theta < 40^{\circ}$, while at large contact angles, the theoretical model based on the thin-sheet equations overestimates the bridge growth. %fails to describe the quantitative behavior of the bridge growth.
For the three-dimensional cases, we find that during coalescence, the cross-section of the bridge is of a spherical cap shape; however, the contact angle of the cross-section is less than the equilibrium contact angle at the initial state, which reveals a non-linear dependence of the growth of the bridge radius and the bridge height at the initial stage. 

%We observe an oscillation at the later stage and the frequency is characterized by lens radius $R$, density $\rho$ and surface tension $\gamma$. 

\section{Method}
We apply the lattice Boltzmann method (LBM) to simulate the dynamics of fluids. In the past decades, the LBM has been used as a powerful tool for numerical simulations of fluid flows~\cite{Succi2001,Kruger2017} and has been extended to simulate multiphase/multicomponent 
fluids~\cite{Shan1993,Cappelli2015}. 
Moreover, the LBM has been applied successfully to investigate viscous and inertial coalescence of suspended droplets~\cite{Gross2013, Lim2017} and sessile droplets on a substrate~\cite{Hessling2017}.
We utilize the pseudopotential multicomponent LBM of Shan and Chen~\cite{Shan1993,Shan1994} with a D3Q19 lattice~\cite{Qian1992} and review some related details in the following.
Three fluid components follow the discretized equation of each distribution function according to the lattice Boltzmann equation
\begin{eqnarray}
  \label{eq:LBG}
 f_i^c(\vec{x} + \vec{e}_i \Delta t , t + \Delta t)=f_i^c(\vec{x},t)+\Omega_i^c(\vec{x},t)
  \mbox{,}
\end{eqnarray}
where $i=1,...,19$. $f_i^c(\vec{x},t)$ are the single-particle distribution functions for fluid component $c=1$, $2$ or $3$, 
$\vec{e}_i$ is the discrete velocity in the $i$th direction, and 
\begin{equation}
  \label{eq:BGK_collision_operator}
  \Omega_i^c(\vec{x},t) = -\frac{f_i^c(\vec{x},t)- f_i^\mathrm{eq}(\rho^c(\vec{x},t), \vec{u}^c(\vec{x},t))}{\left( \tau^c / \Delta t \right)}
\end{equation}
is the Bhatnagar-Gross-Krook (BGK) collision operator~\cite{Bhatnagar1954}. $\tau^c$ is the relaxation time for component $c$. 
Here, $f_i^\mathrm{eq}(\rho^c(\vec{x},t),\vec{u}^c(\vec{x},t))$ is a second-order equilibrium distribution function~\cite{Chen1992}, defined as
\begin{eqnarray}
	\label{eq:eqdis}
	f_i^{\mathrm{eq}}(\rho^c,\mathbf{u}^c) &=& \omega_i \rho^c \bigg[ 1 + \frac{\mathbf{e}_i \cdot \mathbf{u}^c}{c_s^2} \nonumber \\
	&& - \frac{ \left( \mathbf{u}^c \cdot \mathbf{u}^c \right) }{2 c_s^2} + 
	\frac{ \left( \mathbf{e}_i \cdot \mathbf{u}^c \right)^2}{2 c_s^4}  \bigg]
	\mbox{.}
\end{eqnarray}
where $\omega_i$ is a coefficient depending on the direction: $\omega_0=1/3$
for the zero velocity, $\omega_{1,\dots,6}=1/18$ for the six nearest neighbors
and $\omega_{7,\dots,18}=1/36$ for the nearest neighbors in diagonal direction.
$c_s = \frac{1}{\sqrt{3}} \frac{\Delta x}{\Delta t}$ is the speed of sound.
The macroscopic variables, densities and velocities are updated as  
$ \rho^c(\vec{x},t) = \rho_0 \sum_if^c_i(\vec{x},t)$, where $\rho_0$ is a reference density, and $\vec{u}^c(\vec{x},t) = \sum_i  f^c_i(\vec{x},t) \vec{c}_i/\rho^c(\vec{x},t)$, respectively.
\revisedtext{The single relaxation time (SRT) scheme is employed in this work. While the multi-relaxation time (MRT) scheme~\cite{Fan2010} can enhance stability for high Reynolds number or low-viscosity flows, it incurs a higher computational cost due to the required moment-space transformations. Given the low Reynolds numbers and moderate viscosities considered here, SRT provides a stable and computationally efficient alternative.}

Lattice Boltzmann method can be treated as an alternative solver of Navier-Stokes equation in the limit of small Knnudsen and Mach numbers~\cite{Succi2001}.
In our simulations, we choose the lattice constant $\Delta x$, the timestep $ \Delta t$, the reference density $\rho_0 $ and the relaxation time $\tau^c$ to be unity, which results in a kinematic viscosity $\nu^c$ $=$ $\frac{1}{6}$ in lattice units.

In the pseudopotential multicomponent LB method introduced by Shan and Chen, a mean-field interaction force between fluid components $c$ and $c'$ is introduced, written as ~\cite{Shan1993}:
\begin{equation}
  \label{eq:sc}
  \vec{F}_{\mathrm{C}}^c(\vec{x},t) = -\Psi^c(\vec{x},t) \sum_{c'}g_{cc'} \sum_{\vec{x}'} \Psi^{c'}(\vec{x}',t) (\vec{x}'-\vec{x}),\,\,
\end{equation}
in which $\vec{x}'$ denote the nearest neighbours of lattice site $\vec{x}$ and $g_{cc'}$ is a coupling constant determining the surface tension.
$\Psi^c(\mathbf{x},t)$ is called ``effective mass'', defined as~\cite{Shan1993,Shan1994}
\begin{equation}
  \label{eq:psifunc}
  \Psi^c(\vec{x},t) \equiv \Psi(\rho^c(\vec{x},t) ) = 1 - e^{-\rho^c(\vec{x},t)}
   \mbox{.}
\end{equation}
\revisedtext{This effective mass formulation ensures that at low densities it approximates the density $\rho^c$ itself, while at high densities it asymptotically approaches a saturation limit~\cite{CHEN2014210}. The saturation behavior prevents the collapse of the high-density phase, thereby enhancing numerical stability in simulations.}

The force $\vec{F}_{\mathrm{C}}^c(\vec{x},t) $ is then incorporated into the lattice Boltzmann equations by adding a shift $\Delta \vec{u}^c(\vec{x},t) = \frac{\tau^c \vec{F}_{\mathrm{C}}^c(\vec{x},t)}{\rho^c(\vec{x},t)}$ to the velocity $\vec{u}^c(\vec{x},t)$ in the equilibrium distribution function $f_i^\mathrm{eq}$.
The pseudopotential multicomponent LB method is a diffuse interface method with an interface width of $\approx 5\Delta x$, which elegantly removes the stress singularity at the moving contact line usually occurring in sharp-interface models. \revisedtext{We note that in the pseudopotential lattice Boltzmann method, the surface tension is governed by the interaction parameter $g_{cc'}$ in~\eqnref{sc} and is calculated by the Young-Laplace equation. For a spherical droplet of radius $R$ immersed in another fluid, the Young–Laplace equation, $ \gamma=\frac{R\Delta P}{2}$, relates the pressure difference $\Delta P$ over the interface between two fluids to the surface tension $\gamma$.}

We perform simulations in  two and three dimensions. For simplicity and numerical stability, we initialize three fluid components with equal viscosity and density.

\section{Results and discussion}

\subsection{Single liquid lens}
\begin{figure}[bh]
	\includegraphics[width= 0.4\textwidth]{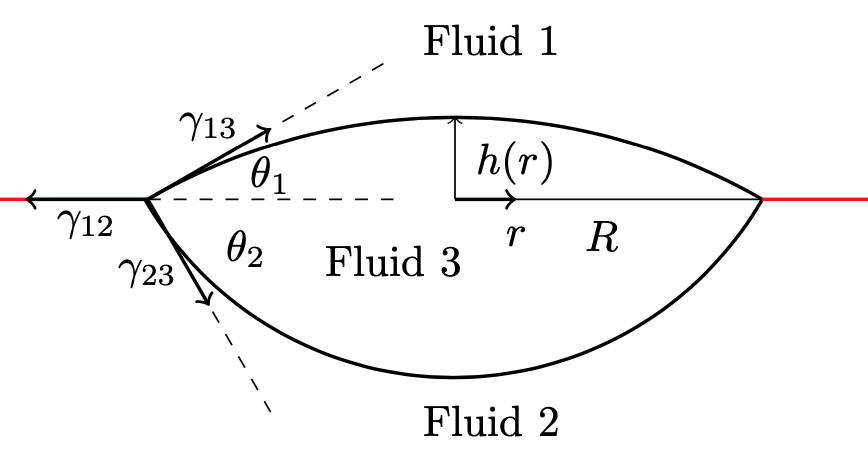}
	\caption{Schematic of a single liquid lens (fluid $3$) at an interface between fluid $1$ and fluid $2$ in the equilibrium state. 
		%$\gamma_{ij}$ is the surface tension at the interface between fluid $i$ and fluid $j$. The base radius of the lens is $R$, and the lens height is $h(r)$ along the radial coordinate.
%The contact angles of the upper and lower parts of the lens are $\theta_1$ and $\theta_2$, respectively.
}
		\label{fig:single-geo}
\end{figure}
\begin{figure*}[ht!]
	\begin{subfigure}{.3\textwidth}
		\includegraphics[width= 0.95\textwidth]{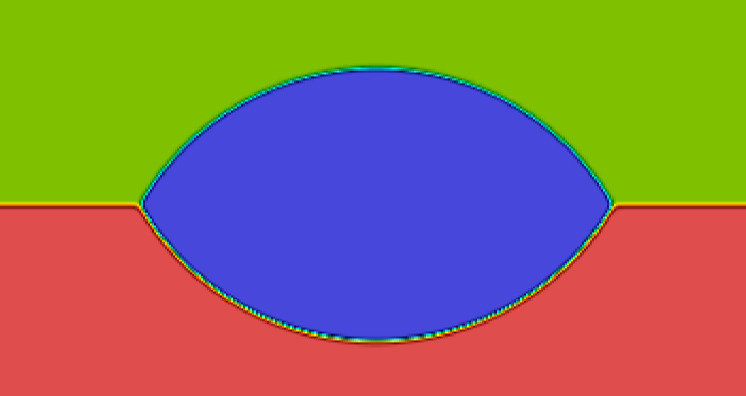}
		\subcaption{$\gamma_{12}:\gamma_{13}:\gamma_{23}= 1:1:1$}
		\label{fig:sim-snapshot1}
	\end{subfigure}
	\begin{subfigure}{.3\textwidth}
		\includegraphics[width= 0.95\textwidth]{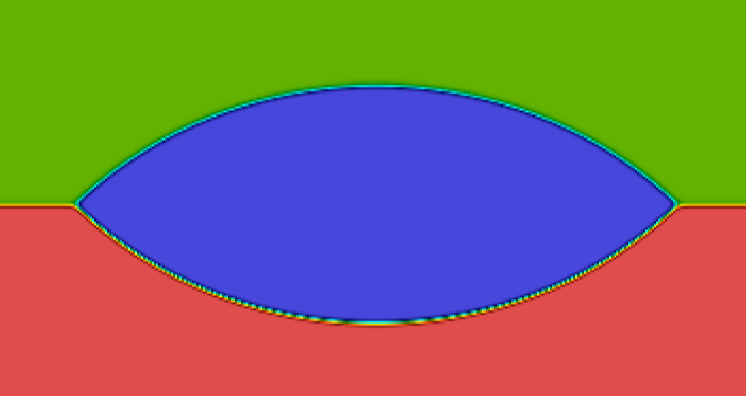}
		\subcaption{$\gamma_{12}:\gamma_{13}:\gamma_{23}= 1.5:1:1$}
		\label{fig:sim-snapshot2}
	\end{subfigure}
	\begin{subfigure}{.3\textwidth}
		\includegraphics[width= 0.95\textwidth]{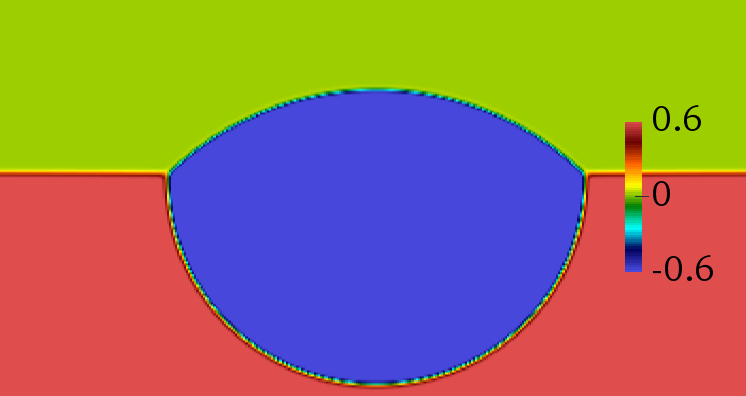}
		\subcaption{$\gamma_{12}:\gamma_{13}:\gamma_{23}= 1:1.5:1$}
		\label{fig:sim-snapshot3}
	\end{subfigure}
	\caption{Snapshots of a single liquid lens in equilibrium state obtained in simulations for different combinations of surface tensions: a) $\gamma_{12}:\gamma_{13}:\gamma_{23}= 1:1:1$ , b) $\gamma_{12}:\gamma_{13}:\gamma_{23}= 1.5:1:1$ , c) $ \gamma_{12}:\gamma_{13}:\gamma_{23}= 1:1.5:1$. The color represents the density difference between lens liquid $2$ and the lower liquid $3$, $\Delta \rho_{23}=\rho_2-\rho_3$.}
	\label{fig:sim-snapshot}
\end{figure*}

We start out to study the spreading of a single liquid lens at a fluid-fluid interface and validate its shape in equilibrium, as illustrated in~\figref{single-geo}. The base radius of the lens is $R$, and the lens height is $h(r)$ along the radial coordinate.
The contact angles of the upper and lower parts of the lens are $\theta_1$ and $\theta_2$, respectively. The equilibrium shape of a liquid lens has been investigated theoretically and experimentally, and has been used as a classical benchmark problem to test the numerical approaches for three-phase flows~\cite{Burton2010, Zheng2019, Yuan2020,Ravazzoli2020}.

When a spherical droplet is initially located at a fluid-fluid interface, it undergoes deformation driven by surface tension to reach its equilibrium state. In the case where gravity is negligible, the liquid lens assumes a spherical cap shape, dominated by surface tension.
At the three-phase contact line, the surface tensions obey the so-called Neumann's law  $ \vec{\gamma}_{12} + \vec{\gamma}_{13} + \vec{\gamma}_{23} = 0$ and after some mathematical manipulations, we obtain  
 \begin{eqnarray}
 \cos \theta_1 = \frac{\gamma^2_{13}-\gamma^2_{23}+\gamma^2_{12} }{2\gamma_{12}\gamma_{13}}\,,  \\ 
  \cos \theta_2 = \frac{\gamma^2_{23}-\gamma^2_{13}+\gamma^2_{12} }{2\gamma_{12}\gamma_{23}}\,,
  \label{eq:neumann}
 \end{eqnarray}
where $\gamma_{ij} = |\vec{\gamma}_{ij}|$.
The height profiles of the upper and lower parts of the lens along the radial coordinate follow
 \begin{align}
h_{up}(r)= \sqrt{(R/\sin \theta_1)^2-r^2} - R\cot \theta_1 \label{eq:height1}, \\ 
h_{lo}(r)= \sqrt{(R/\sin \theta_2)^2-r^2} - R\cot \theta_2. \label{eq:height2}
%h_2=R(1-\cos \theta_2)/\sin\theta_2 
\end{align}

\begin{figure}[h!]
\includegraphics[width= 0.4\textwidth]{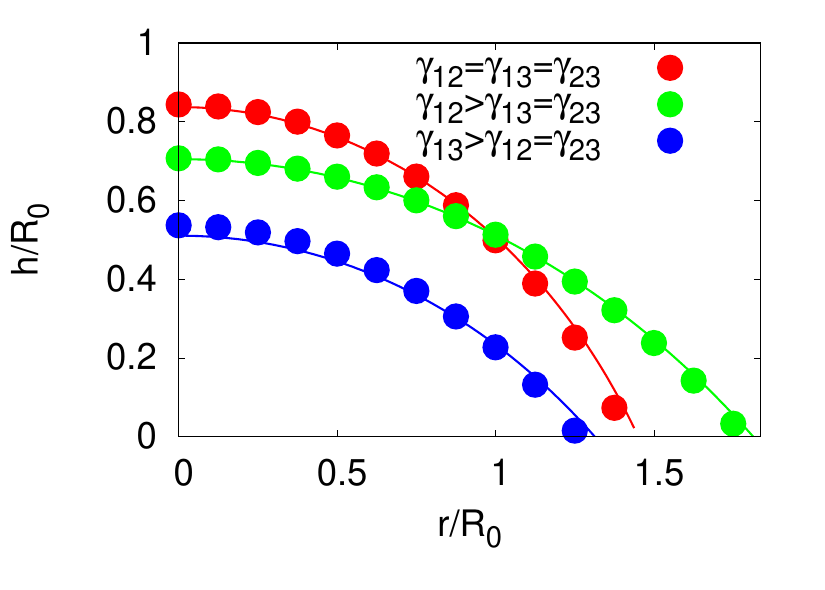}
 \caption{The height profile of the upper half of a single liquid lens in equilibrium state for different surface tension ratios: $\gamma_{12}:\gamma_{13}:\gamma_{23}= 1:1:1$ (red), $\gamma_{12}:\gamma_{13}:\gamma_{23}= 1.5:1:1$ (green)  and $ \gamma_{12}:\gamma_{13}:\gamma_{23}= 1:1.5:1$ (blue). The simulation results (symbols) agree quantitatively with the analytical solution~\eqnref{height1} (solid lines).}
 \label{fig:shape}
\end{figure}

The simulations utilize a 2D system of size
$512 \times 264$. A wall is placed at the bottom of the system, and periodic boundary conditions are applied at the remaining boundaries.  We initialize a spherical droplet of radius $R_0=80$ centered at a fluid-fluid interface and let it equilibrate. The densities of the three fluids are set to $\rho_i= 0.7$. We vary the surface tension ratios to be $\gamma_{12}:\gamma_{13}:\gamma_{23}= 1:1:1$, $\gamma_{12}:\gamma_{13}:\gamma_{23}= 1.5:1:1$  and $ \gamma_{12}:\gamma_{13}:\gamma_{23}= 1:1.5:1$. 
%We initialize a spherical droplet with radius $R_0$ and let it equilibrate, as shown in~\figref{sim-snapshot}. 
%We vary the surface tensions, $\gamma_{12}=\gamma_{13}=\gamma_{23}$~\figref{sim-snapshot1}, 
%$\gamma_{12}=\gamma_{13}=\gamma_{23}$~\figref{sim-snapshot2}, $\gamma_{12}=\gamma_{13}=\gamma_{23}$~\figref{sim-snapshot3}
In the case of $\gamma_{12}:\gamma_{13}:\gamma_{23}= 1:1:1$, the contact angles are 
 $\theta_1 = \theta_2 \sim 60^{\circ}$, as shown in~\figref{sim-snapshot1}. For $\gamma_{12}:\gamma_{13}:\gamma_{23}= 1.5:1:1$, the lens spreads more at the fluid interface to reduce the total surface energy (\figref{sim-snapshot2}), resulting in lower contact angles $\theta_1 = \theta_2 \sim 42^{\circ}$. When $\gamma_{12}:\gamma_{13}:\gamma_{23}= 1:1.5:1$, to minimize the total surface energy, the lens sinks more in the lower fluid (\figref{sim-snapshot3}) and the contact angles are $\theta_1 \sim 42^{\circ}$ and $\theta_2 \sim 97^{\circ}$.
After equilibrium, we measure the height of the liquid lens as a function of the radial coordinate and compare the simulation results with the analytical solution~\eqnref{height1} in~\figref{shape}. We note that determining the exact position of the diffuse interface is challenging, and here we determine the interface position corresponding to zero density difference between the lens liquid and the surrounding liquid, e.g., $\rho^2 - \rho^{3}=0$.
Our simulation results agree quantitatively with the analytical solution~\eqnref{height1}, which demonstrates that our LB model can accurately capture the equilibrium interface shape of a single liquid lens.

\subsection{Two liquid lenses - 2D}
\begin{figure}[t!]
		\includegraphics[width= 0.45\textwidth]{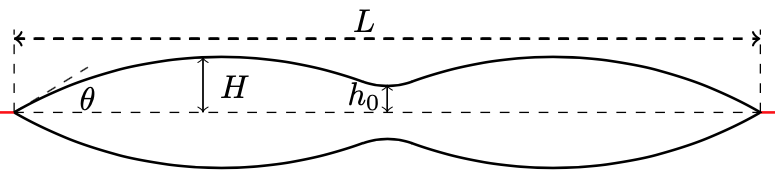}
\caption{Schematic of the side view of two coalescing lenses at a fluid-fluid interface. The lenses are up-down symmetric, and the contact angle is $\theta$. 
The maximal bridge height is $H$, and the bridge height at the center is $h_0$. 
$L$ is the distance between the far ends of the two lenses.}
\label{fig:lenses-geo}
\end{figure}
\begin{figure}[h!]
    \begin{subfigure}{.2\textwidth}
\includegraphics[width= 0.95\textwidth]{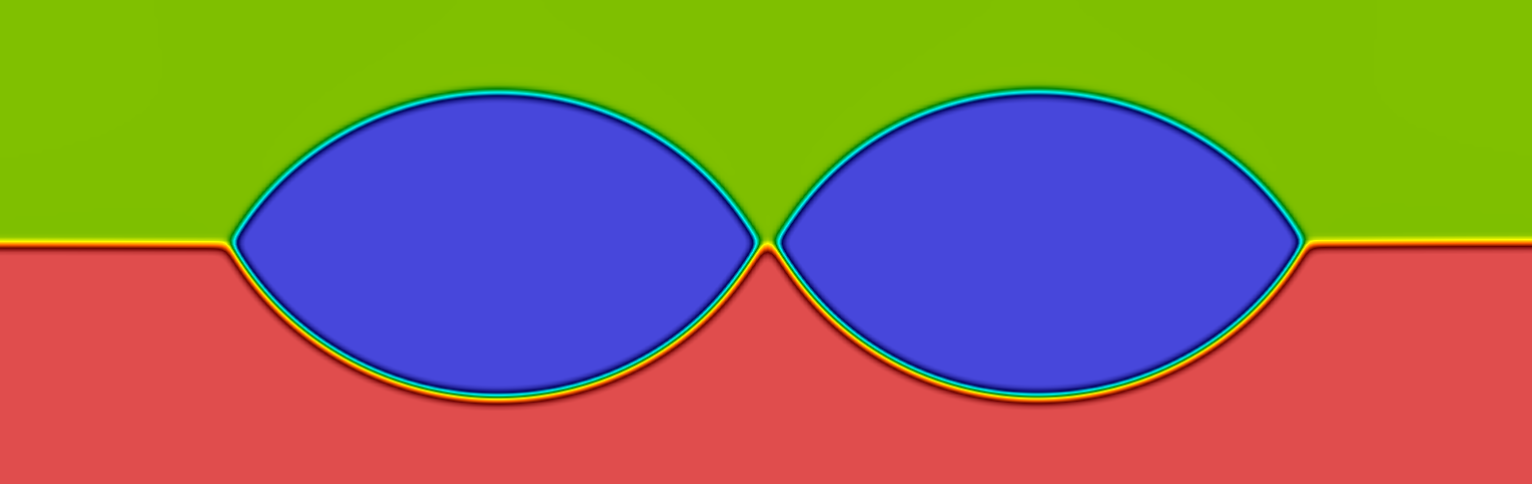}
 \subcaption{}
 \label{fig:2d-1}
 \end{subfigure}
    \begin{subfigure}{.2\textwidth}
\includegraphics[width= 0.95\textwidth]{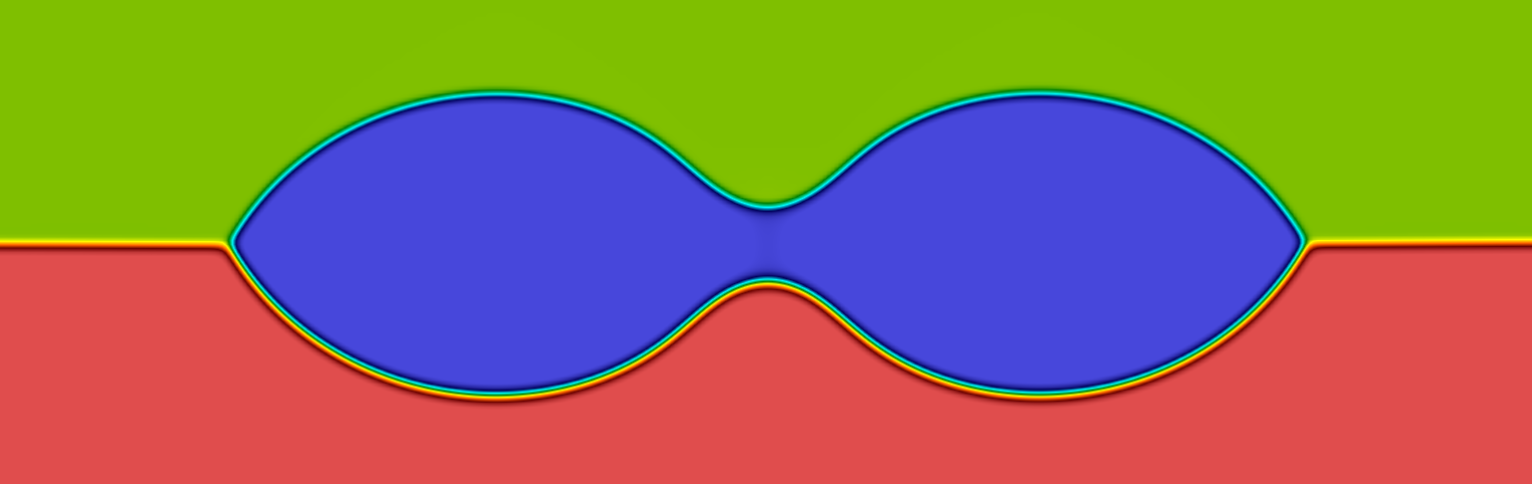}
 \subcaption{}
 \label{fig:2d-2}
 \end{subfigure}
     \begin{subfigure}{.2\textwidth}
\includegraphics[width= 0.95\textwidth]{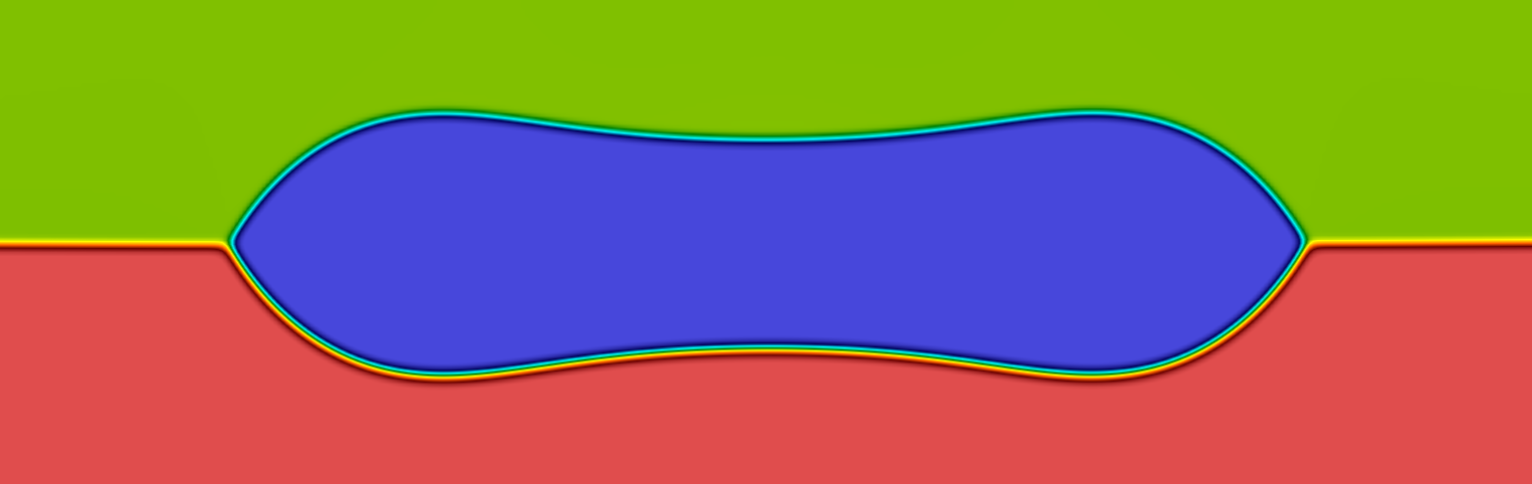}
 \subcaption{}
 \label{fig:2d-3}
 \end{subfigure}
     \begin{subfigure}{.2\textwidth}
\includegraphics[width= 0.95\textwidth]{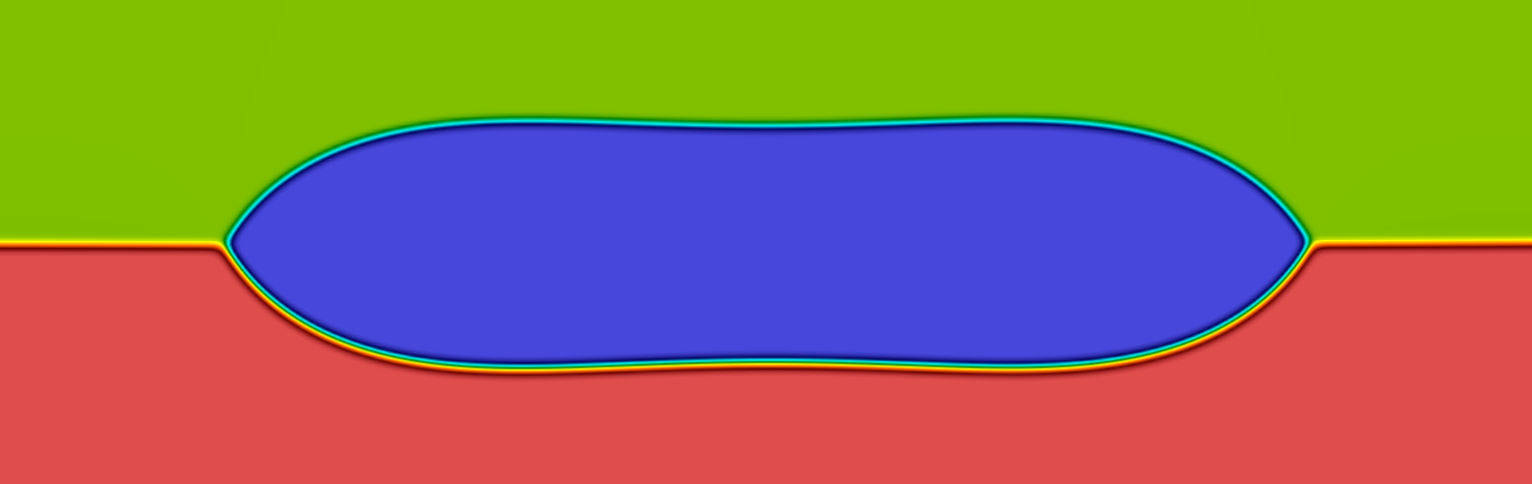}
 \subcaption{}
 \label{fig:2d-4}
 \end{subfigure}
     \begin{subfigure}{.2\textwidth}
\includegraphics[width= 0.95\textwidth]{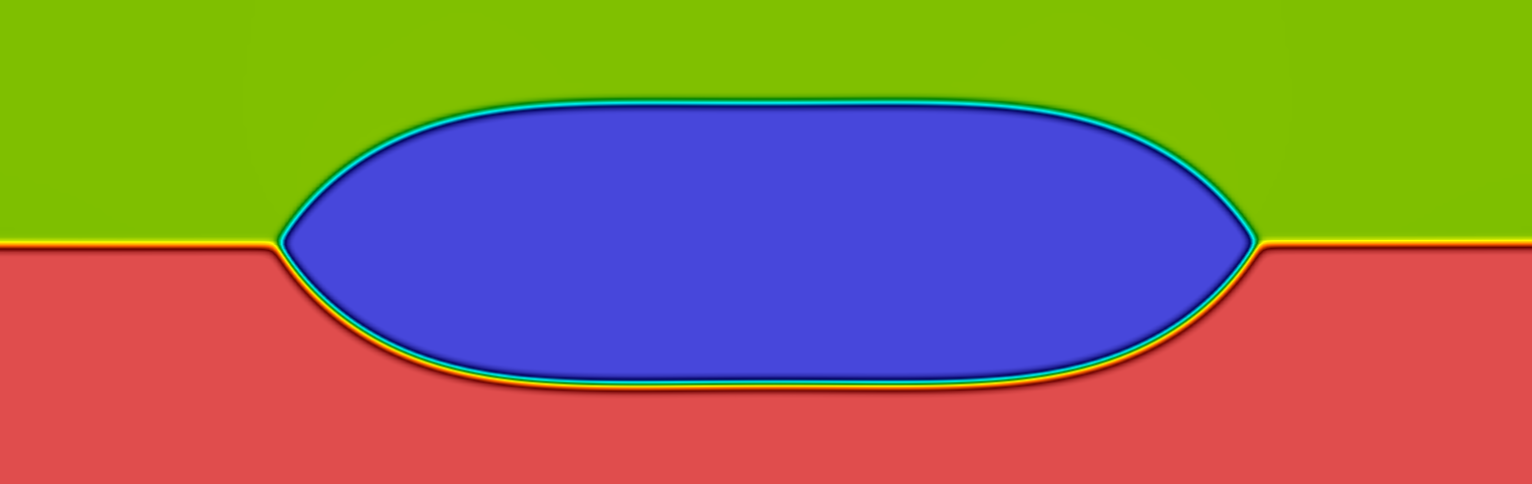}
 \subcaption{}
 \label{fig:2d-5}
 \end{subfigure}
     \begin{subfigure}{.2\textwidth}
\includegraphics[width= 0.95\textwidth]{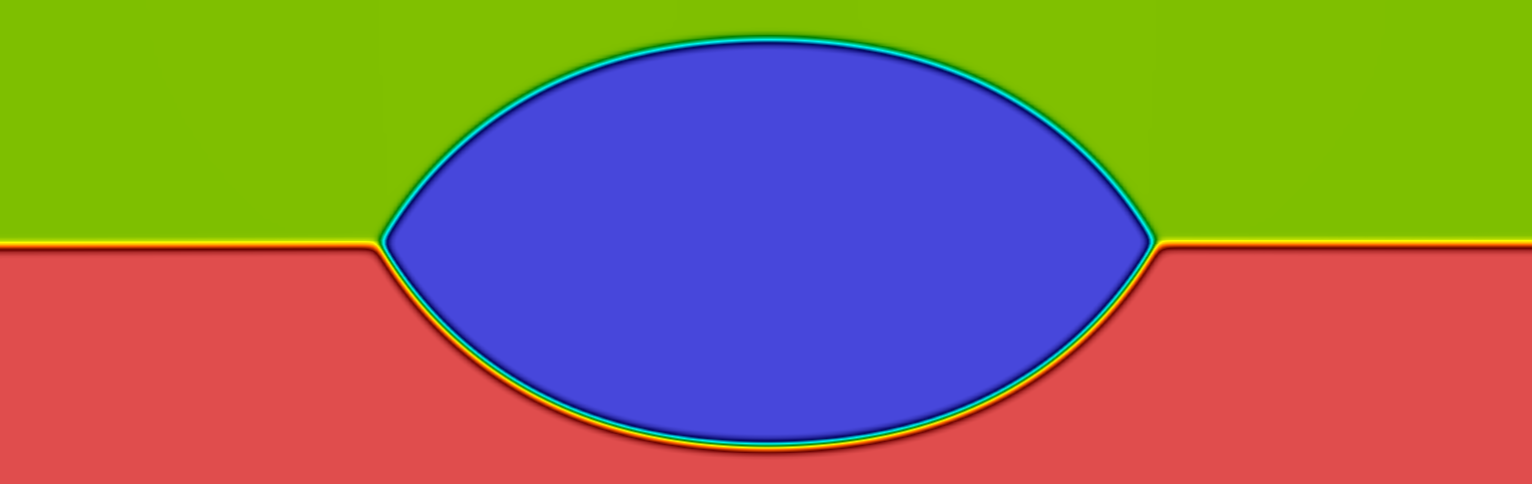}
 \subcaption{}
 \label{fig:2d-6}
 \end{subfigure}
  \caption{Time sequence of the coalescence process of liquid lenses in a 2D system obtained in our simulations. The lenses merge upon contact, connected by a rapidly growing bridge, and ultimately form a larger lens that relaxes into its equilibrium shape. The color represents the density difference between lens liquid $2$ and the lower liquid $3$, $\Delta \rho_{23}=\rho_2-\rho_3$.}
\label{fig:snap-2lenses-2d}
 \end{figure}

 \begin{figure*}[t!]
	\begin{subfigure}{.32\textwidth}
		\includegraphics[width= 0.99\textwidth]{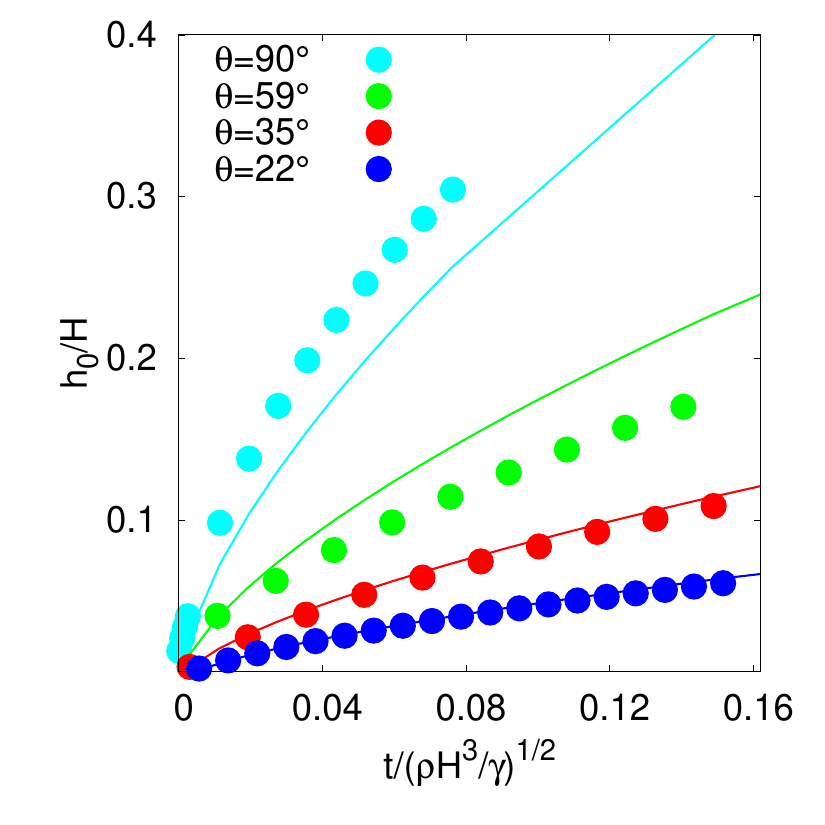}
		\subcaption{}
		\label{fig:h0-t-log}
	\end{subfigure}
	\begin{subfigure}{.32\textwidth}
		\includegraphics[width= 0.99\textwidth]{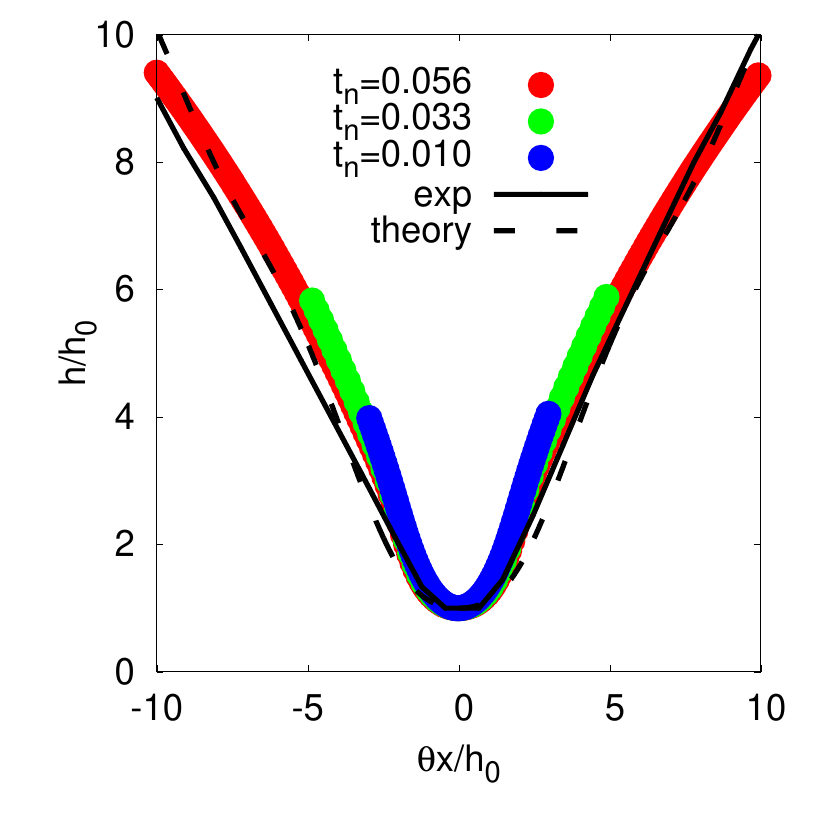}
		\subcaption{}
		\label{fig:height-1}
	\end{subfigure}
	\begin{subfigure}{.32\textwidth}
		\includegraphics[width= 0.99\textwidth]{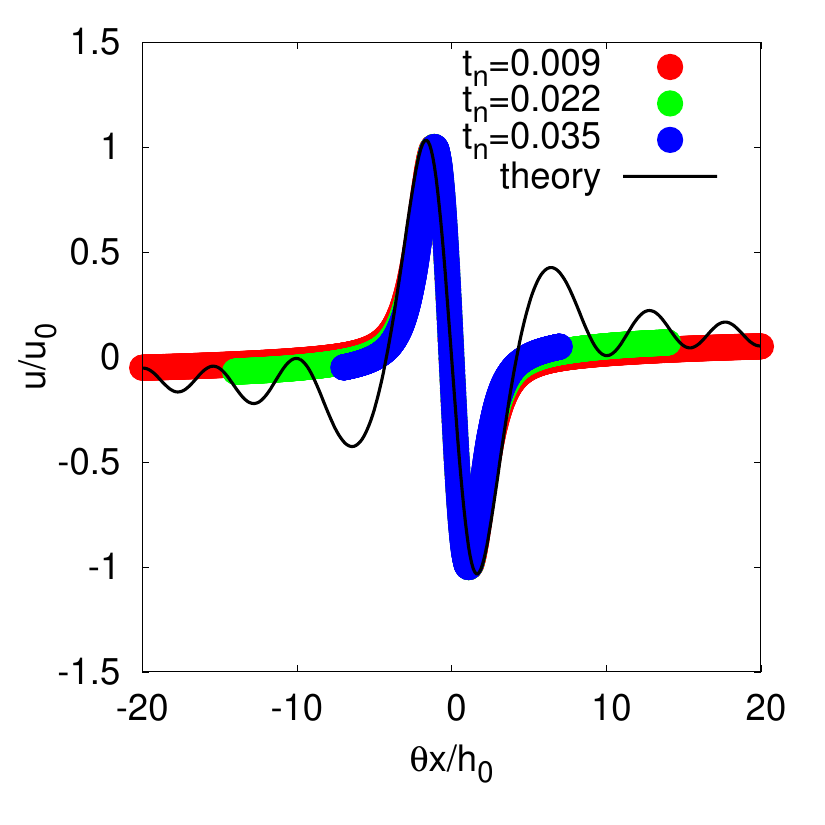}
		\subcaption{}
		\label{fig:velx-1}
	\end{subfigure}
	\caption{
		a) The time evolution of bridge height $h_0$ for different contact angles.  
		The symbols are simulation results and the solid lines correspond to the theoretical prediction~\eqnref{h0-time}. 	
		b) Rescaled bridge profile for lenses with contact angle $\theta=35^{\circ}$ at different times $t_n=t/(\rho H^3/\gamma)^{1/2}$. The collapse of the profiles indicates self-similar dynamics. The symbols represent simulation data, while the solid line represents experimental results~\cite{hack2020}.
		c) Rescaled horizontal velocity profile corresponds to different times $t_n=t/(\rho H^3/\gamma)^{1/2}$. 
		All curves collapse into a single master curve, verifying self-similar dynamics. The solid line is the theoretical prediction~\eqnref{ux-sim}.}
	%the system size is $8960\times1920$.}
	\label{fig:height-velx}
\end{figure*}

Here, we consider the coalescence dynamics of liquid lenses at a fluid-fluid interface, as illustrated in~\figref{lenses-geo}.
The lenses are up-down symmetric, and the contact angles are $\theta_1=\theta_2=\theta$. The maximal bridge height is $H$, and the bridge height at the center is $h_0$. $L$ is the distance between the far ends of the two lenses.
For the coalescence of liquid lenses, the crossover length scale and time scale from the viscous to the inertial regime are determined as~\cite{hack2020} $h_c \sim \frac{\eta^2}{\rho \gamma^2}$ and $t_c \sim \frac{\eta^3}{\rho \gamma^2 \theta^2}$, where $\eta$ is the dynamic viscosity of lens liquid, and $\gamma$ is the surface tension of the lens with respect to its surrounding fluids. 
Using the parameters in our simulations, we obtain $h_c \sim 0.4$ lattice units and $t_c \sim 2$ timesteps, indicating that all our simulations are in the inertial regime.
For liquid lenses with small contact angles, assuming the flow inside the lenses is dominant and parallel to the vertical plane, the dynamics at the initial stage of coalescence are described by 2D thin-sheet equations~\cite{Erneux1993,Scheid2012,Eggers2015,hack2020}, written as 
\begin{eqnarray}
 h_t + (uh)_x &=&0, \\
 \rho (u_t + u u_x) &=& \gamma h_{xxx} + 4\eta \frac{(u_xh)_x}{h} \mbox{,}
 \label{eq:sheet1}
\end{eqnarray}
which represent mass conservation and momentum conservation, respectively. 
Here, $h(x,t)$ describes the shape of the bridge and $u(x,t)$ is the  horizontal velocity of liquid inside the lens. The lower indices denote spatial and temporal derivatives. 

When inertia dominates over viscosity, the above thin-sheet equations are simplified in the inviscid limit to
\begin{eqnarray}
 h_t + (uh)_x &=&0, \label{eq:sheet2h} \\
 \rho (u_t + u u_x) &=& \gamma h_{xxx} \mbox{,}
 \label{eq:sheet2}
\end{eqnarray}
Hack et al.~\cite{hack2020} found experimentally that the bridge growth of two equally sized liquid lenses shows  a self-similar dynamics, which motivated them to solve~\eqnref{sheet2h} and~\eqnref{sheet2} by introducing similarity solutions written as
\begin{equation}
h(x,t) = kt^\alpha \mathcal{H}(\xi), ~~u(x,t) = \frac{\alpha k}{\theta} t^\beta \mathcal{U}(\xi), \label{eq:GeneralAnsatz}
\end{equation}
where $\mathcal H$ and $\mathcal U$ are the similarity functions for the bridge profile and horizontal velocity inside the lenses, and $k$ is a parameter dependent on surface tension, contact angle and density. 
The parameter $\xi= \frac{\theta x}{kt^\alpha}$ is chosen to ensure that $h(x,t) \simeq \theta x$ far away from the bridge.
By inserting~\eqnref{GeneralAnsatz} into ~\eqnref{sheet2h} and~\eqnref{sheet2}, they obtained $\alpha=\frac{2}{3}, \beta=-1/3$. 
By satisfying certain boundary conditions, the time evolution of the bridge height is written as
\begin{equation}
 h_0= kt^{2/3}= \left( \frac{9K_i\gamma \theta^4}{2\rho} \right)^{1/3} t^{2/3},
\label{eq:h0-time}
\end{equation}
in which $K_i=0.106$ is obtained by numerically solving the boundary value problem. The bridge profile and horizontal velocity profile can be written as 
\begin{eqnarray}
% h_0= kt= \left( \frac{9K_i\gamma \theta^4}{2\rho} \right)^{1/3} t^{2/3}, \\
 h(x,t) &=& h_0 \mathcal{H}(\xi), \label{eq:h-sim} \\
u(x,t) &=& \frac{2h_0}{3\theta t} \mathcal{U}(\xi),
\label{eq:ux-sim}
\end{eqnarray}

% \subsubsection{2D}
We conduct simulations of liquid lens coalescence in two dimensions. To capture the initial stage of coalescence where the bridge growth is not affected by the finite height of the lens, a sufficiently large initial height of the lens is required. We initialize a spherical cap-shaped lens at the fluid interface with maximal height $H = 1000$ and contact angles $\theta=22^{\circ},35^{\circ},59^{\circ}$, respectively. \revisedtext{We note that the contact angles were measured after the initially configured single-droplet lens had relaxed to equilibrium.}
% Equal to two cylinder droplets. 
After reaching equilibrium, we numerically mirror the lens and locate it at a distance of $\sim 2$ lattice nodes from the original lens. We note that the initial distance between two lenses has no significant influence on coalescence dynamics~\cite{Hessling2017}, and coalescence is predominantly driven by surface tensions. 
\revisedtext{Given that the lenses approach with negligible relative velocity $v$, the associated Weber number $We=\frac{\rho v^2}{\gamma R}$ is much less than $1$,
which favors coalescence over phenomena like bouncing or breakup. The surface tension during coalescence remains constant, as the system is both isothermal and free of surfactants.}

\figref{snap-2lenses-2d} shows the coalescence of two lenses obtained from our simulations. The color represents the density difference between the lens  and the lower liquid. Upon contact (\figref{2d-1}), a fast-growing bridge is formed connecting the two lenses (\figref{2d-2} and \figref{2d-3}), while the distance $L$ between the far ends of the two lenses remains almost constant. This indicates that at the initial stage of the coalescence, a strong velocity field is located near the bridge center, whereas the influence of coalescence at the far field is negligible. When the bridge height grows to be comparable to the maximal height $H$, the lens retracts (\figref{2d-3}-\figref{2d-5} ), and finally a single lens is formed at the interface which relaxes to its equilibrium shape (\figref{2d-6}).

We show the time evolution of the vertical height of the bridge center $h_0$ in~\figref{h0-t-log} for different contact angles. The bridge grows faster with increasing contact angle. For small contact angles $\theta< 40^{\circ}$, the simulation results (symbols) agree quantitatively with the theoretical analysis~\eqnref{h0-time} (solid lines).
For a large contact angle $\theta=59^{\circ}$, the similarity solution based on the thin-sheet equations overestimates the bridge growth. 
\revisedtext{The thin-sheet equations are based on the lubrication theory, which assumes that the film thickness (normal to the surface) is much smaller than the characteristic length scale along the surface. Consequently, these equations break down at large contact angles, where the interface height violates the underlying thin-film assumption.} We note that our LBM method solves the Navier-Stokes equations and is valid for the whole range of contact angles. In the case of a contact angle of  $\theta=90^{\circ}$, we simulate the coalescence of two suspended spherical droplets in another liquid, and interestingly, we find that the theoretical prediction based on the thin-sheet equations underestimates the bridge growth - contrary to that at a contact angle of $\theta=59^{\circ}$.

\figref{height-1} depicts the bridge profiles of coalescing liquid lenses at a contact angle $\theta = 35^{\circ}$ for different times $t_n=t/(\rho H^3/\gamma)^{1/2}$. We rescale the horizontal coordinate by $\theta/h_0$ and the vertical coordinate by $h_0$, respectively. All the bridge profiles collapse onto a universal curve, verifying the self-similar dynamics at the early stage of coalescence. Our simulation results (symbols) agree quantitatively with experimental results (solid line) and theoretical analysis (dashed line)~\cite{hack2020}. We note that the simulation results slightly deviate from the experimental results at $\theta x/h_0<-3$, possibly because the neighboring lenses are not of exactly equal size in experiments. 
In~\figref{velx-1} we show the horizontal velocity profiles $u$ at different times $t_n=t/(\rho H^3/\gamma)^{1/2}$ near the bridge center. We rescale the horizontal coordinate with $\theta/h_0$ and the vertical coordinate with the maximal horizontal velocity $u_0(t_n)$ at its corresponding time $t_n$. We find that the velocity profiles $u_x$ also collapse to a single curve, indicating self-similar dynamics. In contrast to the theoretical prediction~\eqnref{ux-sim} (solid line), our simulations do not exhibit strong oscillations of the horizontal velocity, which is likely due to the presence of viscous damping in our simulation. In contrast, the theoretical analysis assumes an inviscid limit.

\subsection{Two liquid lenses - 3D}
Here, we carry out simulations of liquid lens coalescence in three dimensions. 
\begin{figure}[h!]
	\begin{subfigure}{.15\textwidth}
		\includegraphics[width= 0.9\textwidth]{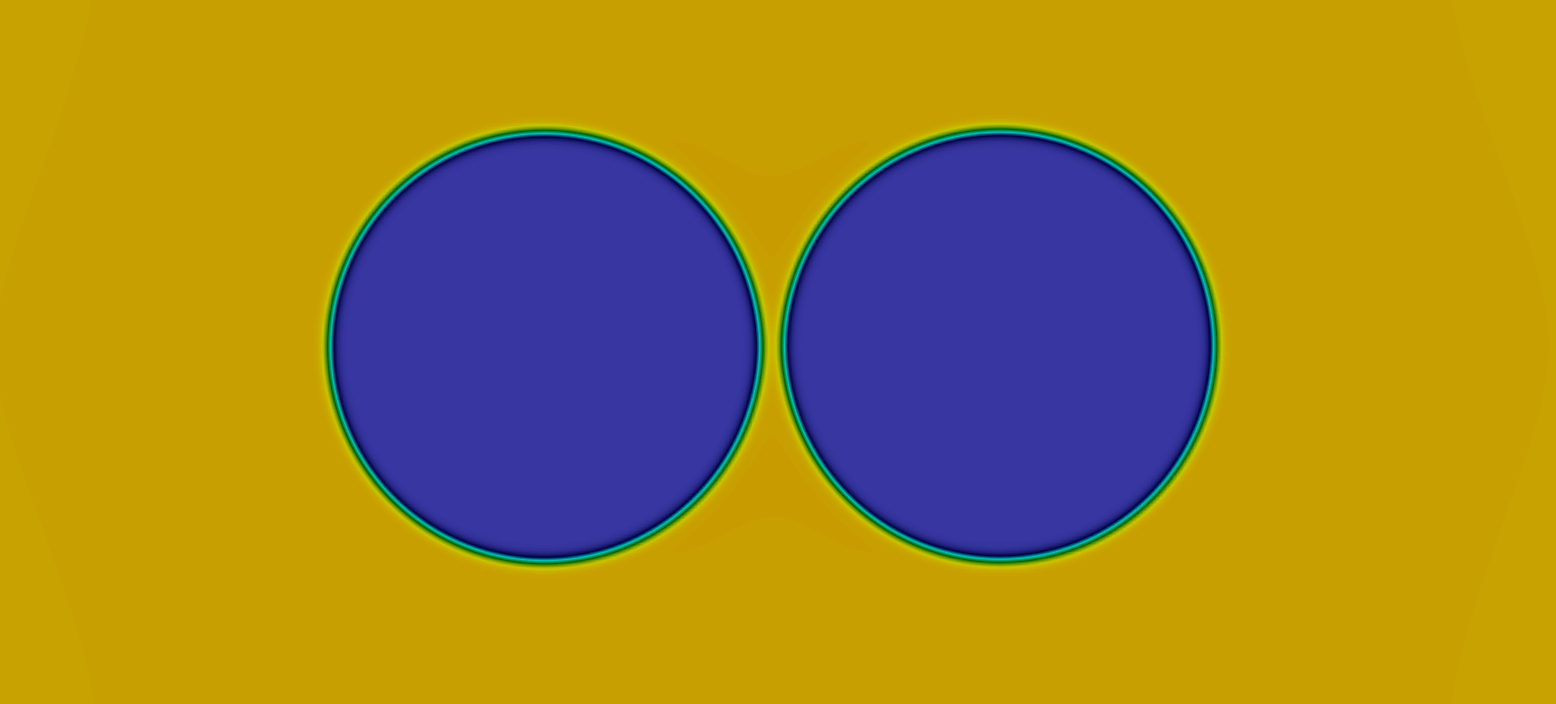}
		\subcaption{}
		\label{fig:3d-1}
	\end{subfigure}
	\begin{subfigure}{.15\textwidth}
		\includegraphics[width= 0.9\textwidth]{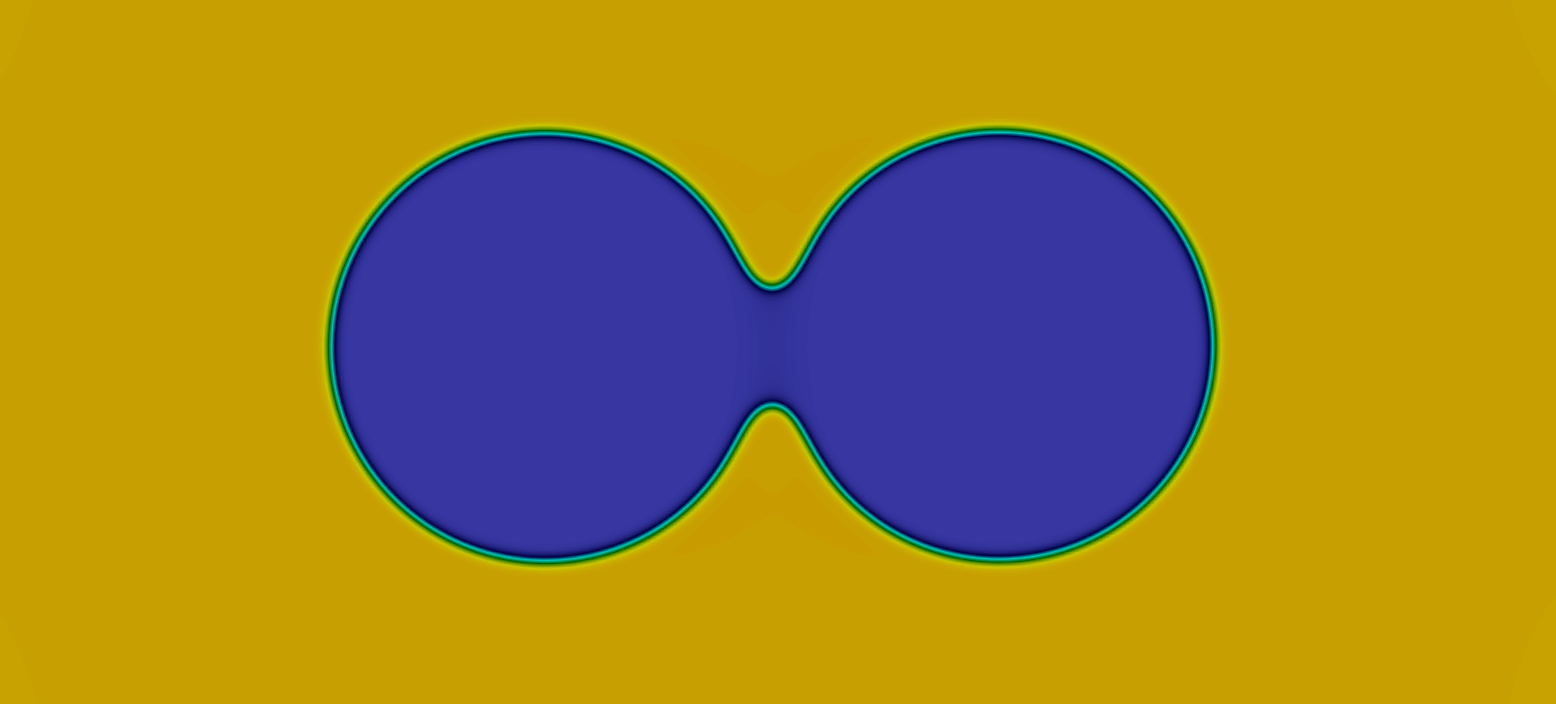}
		\subcaption{}
		\label{fig:3d-2}
	\end{subfigure}
	\begin{subfigure}{.15\textwidth}
		\includegraphics[width= 0.9\textwidth]{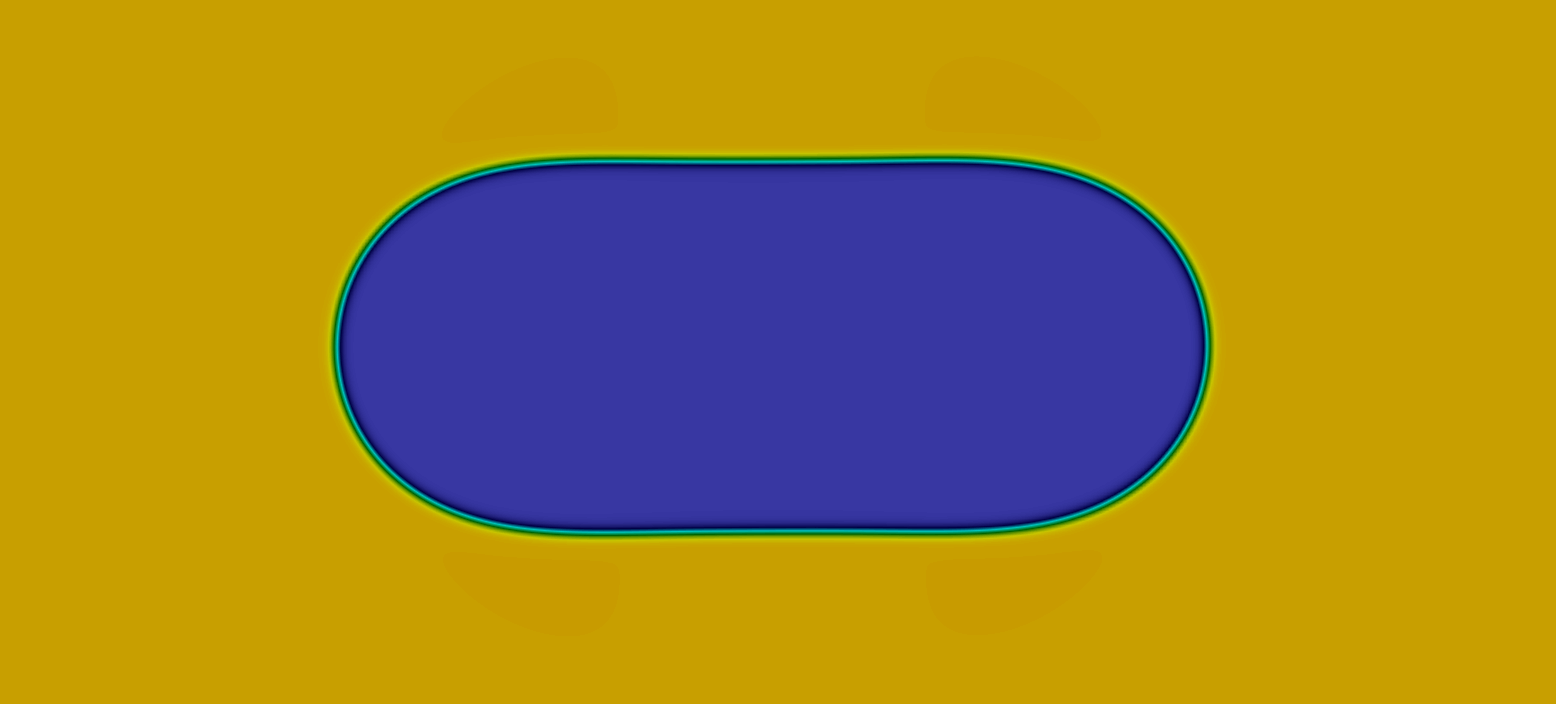}
		\subcaption{}
		\label{fig:3d-3}
	\end{subfigure}
	\begin{subfigure}{.15\textwidth}
		\includegraphics[width= 0.9\textwidth]{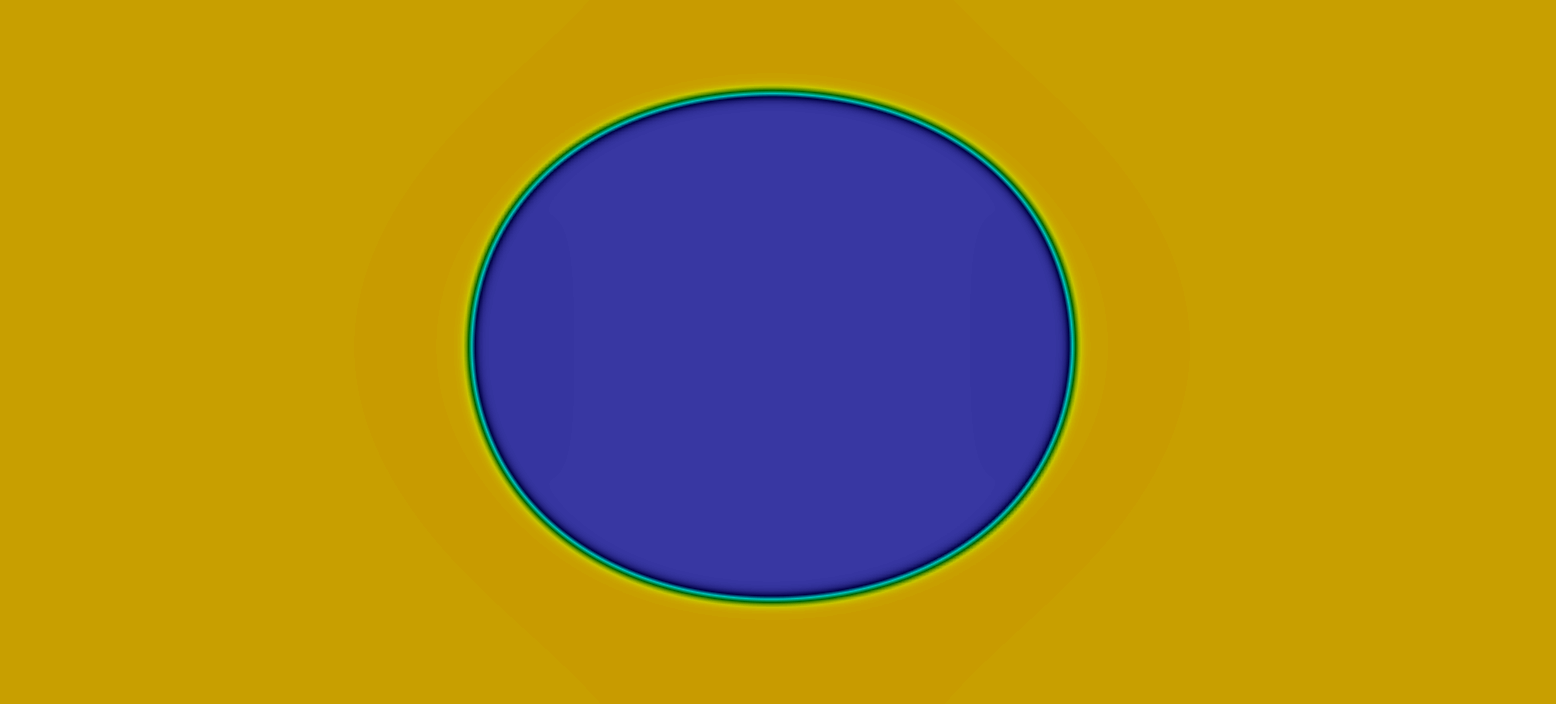}
		\subcaption{}
		\label{fig:3d-4}
	\end{subfigure}
	\begin{subfigure}{.15\textwidth}
		\includegraphics[width= 0.9\textwidth]{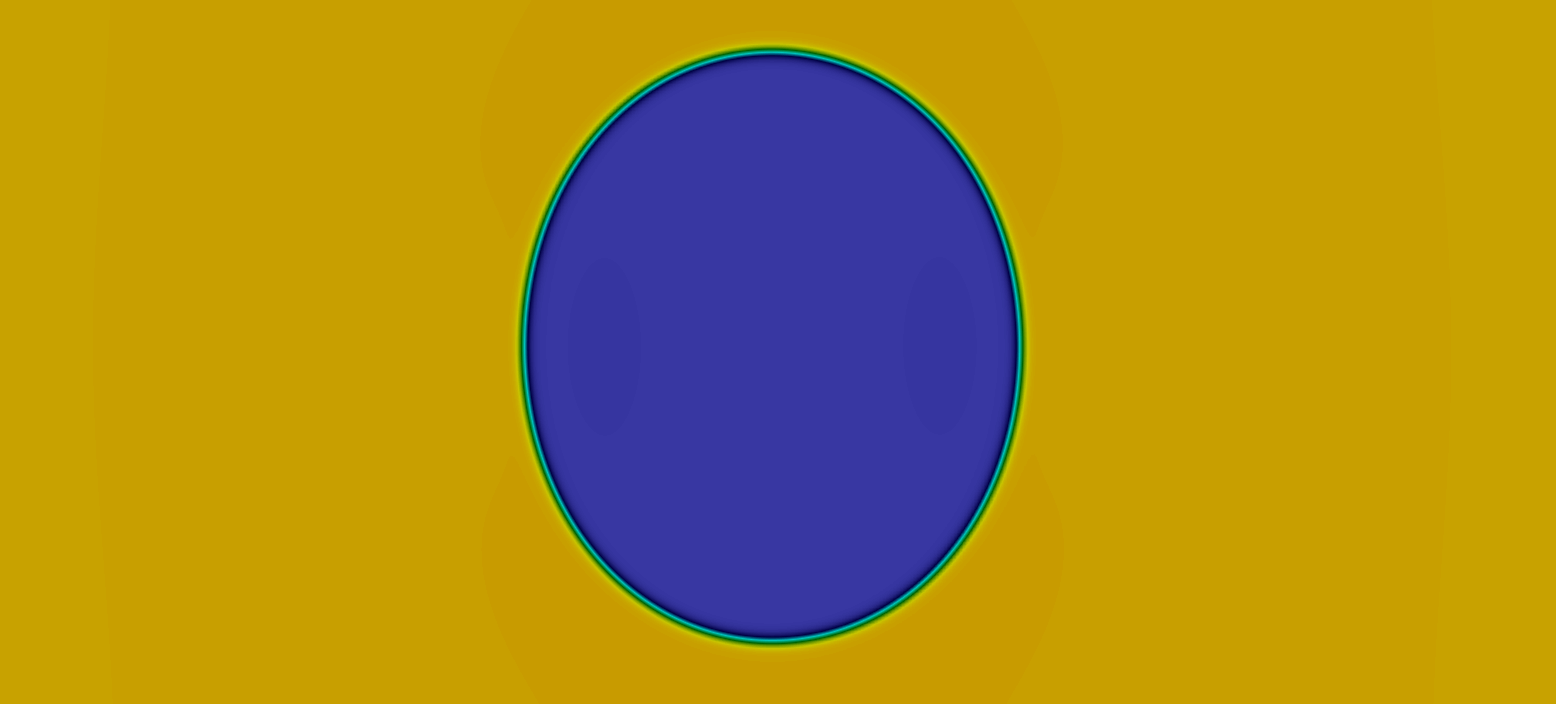}
		\subcaption{}
		\label{fig:3d-5}
	\end{subfigure}
	\begin{subfigure}{.15\textwidth}
		\includegraphics[width= 0.9\textwidth]{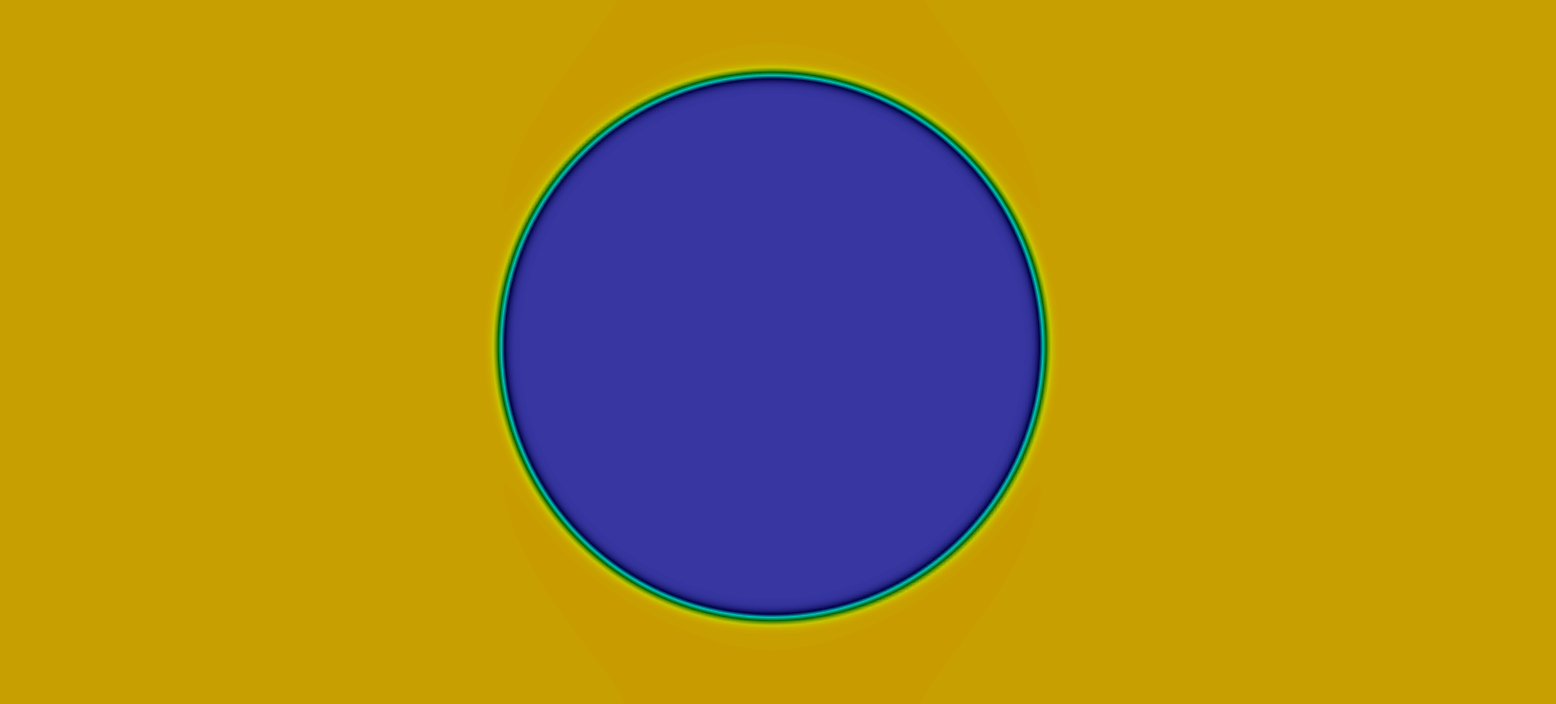}
		\subcaption{}
		\label{fig:3d-6}
	\end{subfigure}
	\caption{Time sequence of the top-views of two coalescing lenses in three dimensions.}
	\label{fig:snap-2lenses-3d}
\end{figure}

\begin{figure*}[t!]
	\begin{subfigure}{.32\textwidth}
		\includegraphics[width= 0.99\textwidth]{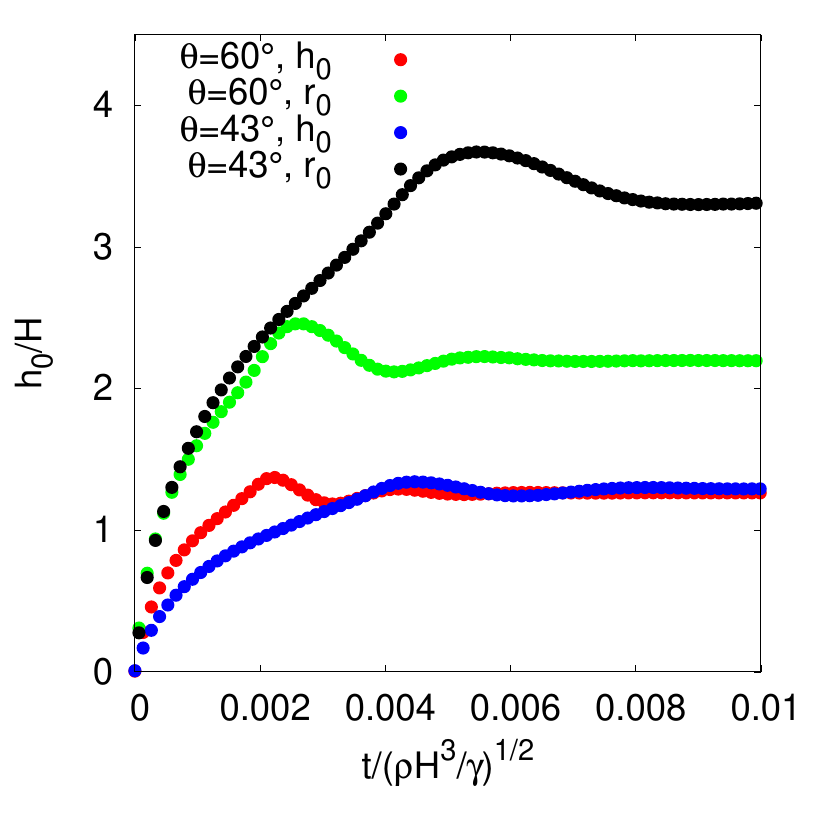}
		\subcaption{}
		\label{fig:height-3d}
	\end{subfigure}
	\begin{subfigure}{.32\textwidth}
		\includegraphics[width= 0.99\textwidth]{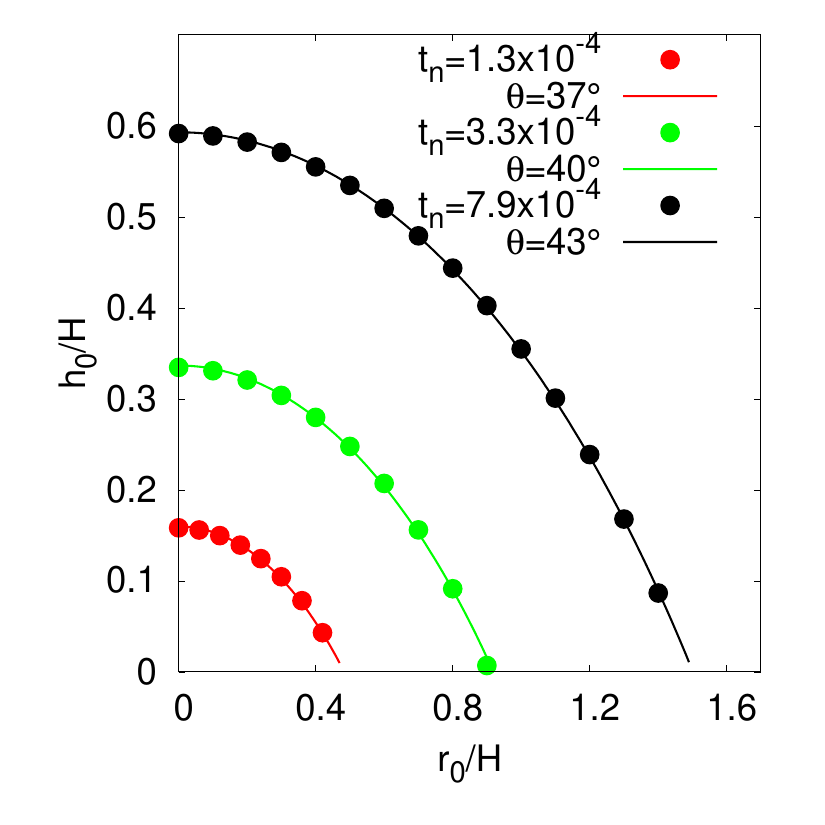}
		\subcaption{}
		\label{fig:shape-3d}
	\end{subfigure}
	\hspace{1mm}
	\begin{subfigure}{.32\textwidth}
		\includegraphics[width= 0.99\textwidth]{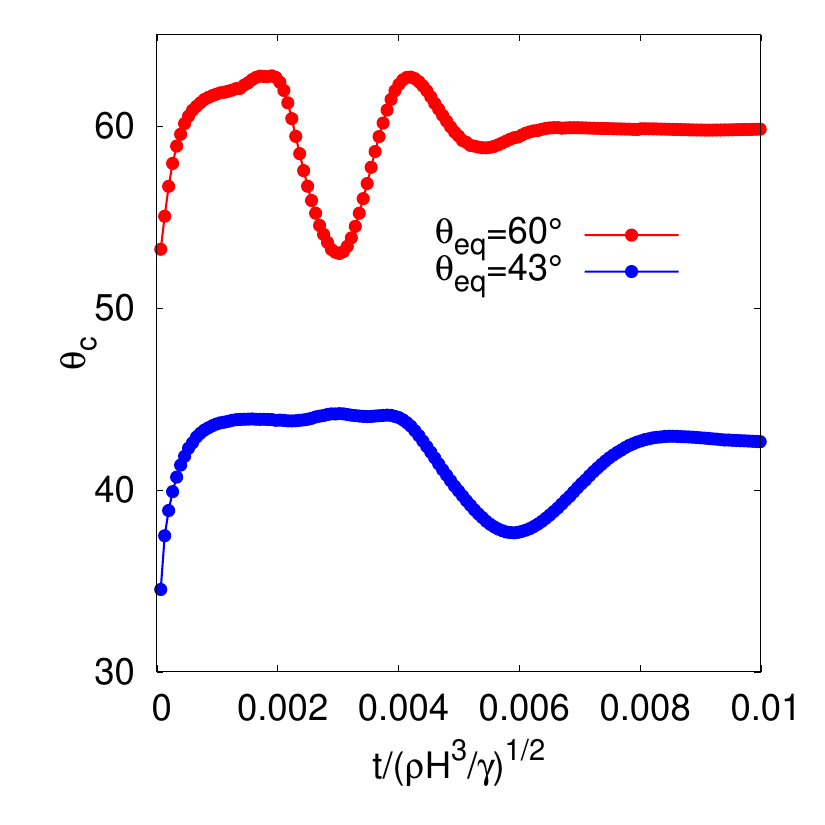}
		\subcaption{}
		\label{fig:theta-3d}
	\end{subfigure}
	\caption{a) the time evolution of bridge height $h_0$ and radius $r_0$ for different contact angles $\theta=60^{\circ}$ and $\theta=43^{\circ}$, respectively.
	b) the shape of the cross-section of the bridge at different times. c) the time evolution of contact angle $\theta_c$ of the bridge cross section. }
	\label{fig:height-radius-3d}
\end{figure*}
The computational cost is extremely high if we simulate lenses with an initial height $H\sim 1000$. Therefore, to reduce computational cost, 
we initialize a spherical cap-shaped lens at the fluid interface with maximal height $H = 100$ and contact 
angles $\theta=43^{\circ},60^{\circ}$, respectively.
We let the system equilibrate, then we mirror it to locate it next to the initial lens at a distance of $\sim 2$ lattice units.
\figref{snap-2lenses-3d} shows the top-view of the coalescence process of two lenses obtained 
in our simulations. Similar to the side-views of the 2D simulations (\figref{snap-2lenses-2d}), the bridge connecting two lenses grows 
fast initially (\figref{3d-1}, (\figref{3d-2} and \figref{3d-3}). At a later stage, 
the lens retracts (\figref{3d-3}-\figref{3d-5}), followed by oscillations and finally relaxes to its equilibrium shape (\figref{3d-6}). 

In~\figref{height-3d}, we show the time evolution of the bridge height $h_0$ and 
radius $r_0$ for different contact angles $\theta=60^{\circ}$, $43^{\circ}$, respectively. 
The vertical height of the bridge grows faster with a larger contact angle, similar to what was observed in the $2D$ simulations (see \figref{h0-t-log}).
Interestingly, the horizontal radius of the bridge grows at the same speed for both contact angles in the initial stage, which indicates that 
the bridge radius growth is not affected by the initial contact angle of the lenses, and the growth of the bridge radius is not a linear function of the growth of bridge height, in contrast
to that observed in coalescing sessile droplets on a substrate. 
A possible explanation is that due to the no-slip boundary condition, the coalescence speed of sessile droplets on a substrate is relatively slow, 
giving sufficient time for the cross-section of the bridge to relax immediately to a spherical cap shape with a corresponding equilibrium contact angle.
However, the coalescence speed of liquid lenses at a fluid-fluid interface is significantly faster, and the bridge height and radius grow in a non-coupled manner, resulting in a non-linear dependence of bridge radius and height growth. 
\revisedtext{Although our simulations were limited to contact angles of $\theta = 43^\circ$ and $60^\circ$ due to computational resource constraints, the initial growth of the bridge radius is expected to be independent of the contact angle across a wider range. This expectation is based on previous work, which shows that 
the growth of the width of coalescing lenses agrees with studies for freely suspended, respectively spherical droplets~\cite{Scheel2023}. 
If one considers a horizontal cross-section through the liquid lenses, their coalescence resembles that of freely suspended droplets with an effective contact angle of $90^{\circ}$. In this view, the contact angle observed in the vertical cross-section plays a negligible role in the initial stage.}

Next, we explore the shape of the cross-section of the bridge. \figref{shape-3d} depicts the upper quarter of the lens for different times $t_n$ 
obtained in our simulations (symbols). The shape of the cross-section of the bridge can be accurately described 
by a spherical cap (solid lines), dominated by surface tensions. 
However, the contact angles of the cross-section obtained by fitting the shape with a spherical cap function vary over time. 
\figref{theta-3d} shows the time evolution of the contact angle $\theta_c$ of the cross section of the bridge. 
The contact angle $\theta_c$ is less than the equilibrium contact angle $\theta_{eq}$ and 
increases rapidly in the initial stage, which indicates that the bridge radius grows faster than the bridge height.
The contact angle reaches a plateau at the intermediate stage, demonstrating a linear relation of the growth of bridge height and bridge width, $r_0/h_0 = \cos\theta_c/(1-\sin\theta_c)$. Afterwards, due to oscillation of the lens, the contact angle decreases, followed by an increase. 
Finally, the contact angle $\theta_c$ of the cross-section of the bridge arrives at the equilibrium contact angle.

\section{Conclusion}

We numerically investigate the inertial coalescence of liquid lenses over a wide range of contact angles using the pseudopotential lattice Boltzmann method.
\revisedtext{While prior work has focused largely on liquid lenses with small contact angles
and only addresses the initial stage of coalescence, our work extends the analysis to the coalescence of liquid lenses with large contact angles and the coalescence dynamics at later stages.}

Our simulations in two dimensions successfully capture the self-similar dynamics of both the bridge and velocity profiles, showing quantitative agreement with experimental observations and a theoretical framework based on thin-sheet equations.
Furthermore, our simulation results demonstrate that the thin-sheet equations are applicable to describe the quantitative behavior of bridge growth for small contact angles approximately up to $\theta < 40^{\circ}$.
In the three-dimensional case, we find out that the bridge height grows faster with a larger contact angle, while the growth of the bridge radius is independent of the contact angle at the initial stage. The contact angle of the cross-section of the bridge increases rapidly in the initial stage. It reaches a plateau at the intermediate stage, indicating a transition from a non-linear to a linear dependency between the growth of bridge height and bridge width. \revisedtext{A theoretical model capable of predicting the transition's critical conditions—such as the critical bridge radius or time—and describing the full bridge growth dynamics is not yet available. Developing such a model extends beyond the scope of the current work and represents a valuable direction for future research.}

The pseudopotential lattice Boltzmann method has been widely applied to study binary fluid component systems~\cite{Kruger2017,liu_multiphase_2016,xie_effect_2025}, and our results demonstrate its applicability to ternary fluid systems, potentially inspiring its use for investigating printing and coating multi-component solutions for functional material synthesis in catalytic and electronic applications~\cite{vinodh_recent_2024,steinberger_challenges_2024}.
\revisedtext{For instance, in wet-on-wet inkjet printing, the final deposition pattern is affected by the competition between coalescence and evaporation timescales~\cite{Wijshoff2018}. Our findings provide an estimate of the coalescence timescale, thereby offering practical guidelines for optimizing drying conditions.}

\begin{acknowledgements}
We thank Jacco Snoeijer, Michiel Hack, and Walter Tewes for fruitful discussions.
Financial support is acknowledged from the Deutsche
Forschungsgemeinschaft (DFG, German Research Foundation) Project-ID 431791331 (CRC1452 CLINT), and the German Federal Ministry of Education and Research (BMBF) -- Project H2Giga/AEM-Direkt (Grant number 03HY103HF).
We thank the Gauss Centre for Supercomputing e.V.
(\url{www.gauss-centre.eu}) for funding this project by providing computing time
through the John von Neumann Institute for Computing (NIC) on the GCS
Supercomputer JUWELS at Jülich Supercomputing Centre (JSC).
\end{acknowledgements}
\bibliographystyle{unsrt}
 \bibliography{biblio/thesis-ref.bib}
\end{document}